\documentclass[usegraphicx]{mn2e}

\usepackage{times, amssymb}

\title[The instability strip of post-AGB stars]{Defining the instability 
strip of pulsating post-AGB binary stars from ASAS and NSVS photometry}
\author[Kiss et al.]{L. L. Kiss$^1$, A. Derekas$^1$, Gy. M. Szab\'o$^2$\thanks{Magyary Zolt\'an 
Postdoctoral Research Fellow}, T. R. Bedding$^1$, L. Szabados$^3$\\
\\
$^1$School of Physics A28, University of Sydney, NSW 2006, Australia\\
$^2$Department of Experimental Physics and Astronomical Observatory, University of
Szeged, Hungary\\
$^3$Konkoly Observatory of the Hungarian Academy of Sciences, Budapest, Hungary}

\begin{document}

\date{Accepted ... Received ..; in original form ..}


\maketitle

\begin{abstract}

We analyse public domain time-series photometric observations of 30 known and candidate
binary post-AGB stars for measuring pulsation and orbital periods. We derive
periodicities for 17 stars for the first time in the literature. Besides identifying five
new RV~Tauri type pulsating variables (three with the RVb phenomenon, i.e. long-term
changes of the mean brightness), we find multiply periodic (or possibly irregular)
post-AGB stars on the two edges of the instability strip. The temperature
dependence of the peak-to-peak light curve amplitudes clearly indicates the changes in
excitation as post-AGB stars evolve through the strip. One object, the peculiar Type II
Cepheid ST~Pup, showed a period increase from 18.5 to 19.2~d, which is consistent with
the known period fluctuations in the past. In HD~44179, the central star of the
Red Rectangle nebula, we see very similar asymmetric light curve than was measured 10--15
years ago, suggesting a very stable circumstellar environment. In contrast to this,
HD~213985 shows coherent but highly non-repetitive brightness modulation, indicating
changes in the circumstellar cloud on a similar time-scale to the orbital period.  

\end{abstract}


\section{Introduction}

Post-AGB stars are rapidly evolving descendants of low and intermediate mass stars
($\lesssim$8 M$_\odot$). They are evolving at constant luminosity, crossing the
Hertzsprung--Russell diagram from the cool Asymptotic Giant Branch (AGB) to the ionizing
temperature of the planetary nebula nuclei, on a time-scale of about  10$^4$ years. During
this process they cross the classical instability strip, in which large-amplitude
radial oscillations are driven by the $\kappa$-mechanism, and the objects will be
recognized as Type II Cepheids and RV~Tauri type pulsating variables. Because of their
fast evolution, these stars are rare and not many have been identified in the Galaxy
(Van Winckel 2003).

The use of pulsations to determine the physical parameters of post-AGB stars has been
very limited so far. Classical post-AGB variables, like (longer period) Type II Cepheids
and RV~Tauri stars, have an extensive literature and are understood to be  low-order
radial pulsators (for a review, see Wallerstein 2002). Some RV~Tau stars, the so-called
RVb variables, also show long-term changes that are attributed to binarity. Although the
MACHO database yielded important results on these objects (Alcock et al. 1998), the full
use of microlensing data of post-AGB stars is yet to be explored. The situation is much
worse for post-AGB pulsators outside the Cepheid instability strip. Theoretical models
have suggested that complex, possibly  chaotic, pulsations should occur in stars hotter
than the instability strip (Aikawa 1991, 1993; Gautschy 1993; Zalewski 1993). Indeed,
irregular pulsations were found in several cases (Fokin et al. 2001; Le Coroller et al.
2003), but a comprehensive survey of many objects, using a homogeneous set of long-term
observational data, has rarely been done so far (e.g. Pollard et al. 1996, 1997). This
group of stars not only offer the possibility of measuring  physical parameters via
modelling oscillations, but they also offer the best opportunity to observe stellar
evolution over human time-scales. Because these stars evolve quickly across the HRD,
changes of their pulsation periods may be a sensitive indicator of the ongoing
evolution, as in the extreme example of FG~Sagittae (Jurcsik 1993).

De Ruyter et al. (2006) recently analysed broad-band Spectral Energy Distribution (SED)
characteristics of a sample of post-AGB objects to investigate the presence of Keplerian
rotating passive discs. The sample contained 51 stars, including confirmed binary
post-AGB stars collected from the literature, classical RV~Tauri stars from the General
Catalog of Variable Stars (GCVS, Kholopov et al. 1985-1988) with strong infrared excess
and a new sample of candidate RV~Tauri stars (Lloyd Evans 1997). Based on the very
similar SEDs they concluded that in all systems, gravitationally bound dusty discs are
present, implying binarity of the central object. De Ruyter et al. (2006) used
pulsations to estimate luminosities via the RV~Tauri period--luminosity relation (Alcock
et al. 1998). However, they only found pulsation periods in the literature for 21 stars,
even though a much higher fraction of their sample resides in the instability strip.
This lack of information prompted us to examine the  publicly available photometric 
databases. Here we present our findings on photometric variability of the De Ruyter et
al. sample.

\section{The sample and data analysis}

\begin{figure*}
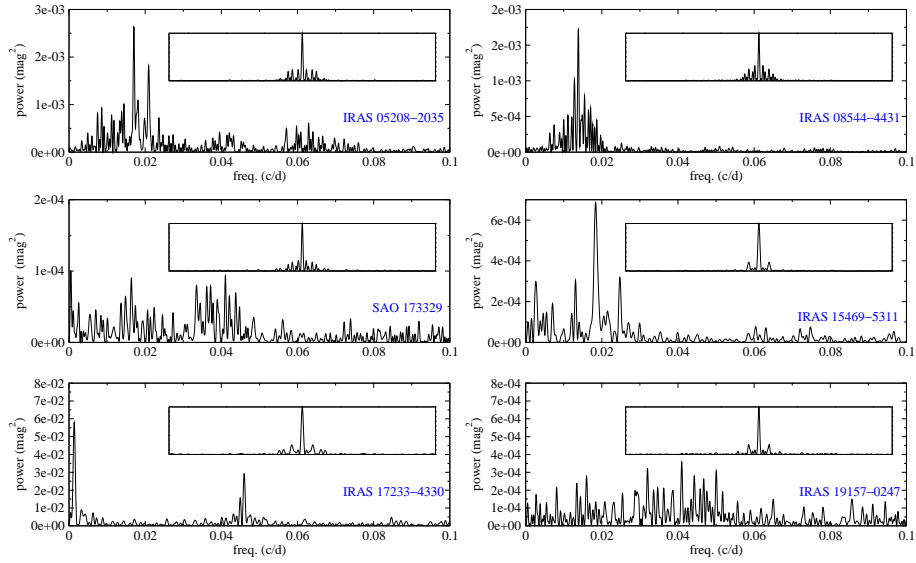

\begin{center}  
\leavevmode
\includegraphics[width=6cm]{fig1a.eps}
\includegraphics[width=6cm]{fig1b.eps}
\includegraphics[width=6cm]{fig1c.eps}
\includegraphics[width=6cm]{fig1d.eps}
\end{center}
\caption{Sample power spectra. The insets show the spectral
window on the same scale.}
\label{power}
\end{figure*}

We used two all-sky photometric surveys as light curve sources: ASAS and NSVS. The third
upgrade of All Sky Automated Survey (ASAS) has been covering 70\% of the sky (everything
south of $+28^\circ$ declination) since 2000. As of writing, the ASAS-3 system has
produced  a few hundred individual $V$-band magnitude measurements for each of about 10
million stars,  using two wide-field telescopes (200 mm telephoto lenses), located at
the Las Campanas Observatory  (see Pojmanski 2002 and references therein for details).
The data have been online since the start of the observations and are available for
download through the ASAS-webpage\footnote{http://archive.princeton.edu/$\sim$asas}. The
useful magnitude range is between $V=8-14$ mag, although in some cases even brighter
variables (e.g. U~Mon) have good light curves, presumably due to shorter exposure times
in these fields. The typical photometric errors (as indicated by the point-to-point
scatter of long period variables) range from 0.01 to 0.2--0.4 mag, depending on the
target brightness.

A limited northern counterpart to the ASAS was the Northern Sky Variability Survey
(NSVS), which captured 1 year of the northern sky over the optical magnitude range from
8 to 15.5, providing typically 100--500 unfiltered measurements for approximately 14
million objects. The instrument was a wide-field telescope similar to that of the ASAS
project and took data between 1999 April 1 and 2000 March 30 in Los Alamos, New
Mexico (Wo\.zniak et al. 2004). The NSVS public data release is available through a
dedicated webpage\footnote{http://skydot.lanl.gov/nsvs/nsvs.php}.

We searched the two online databases for the 30 stars that have no pulsation period given
in Table\ A.2. of De Ruyter et al.  (2006). We found useful data for 22 (mostly southern)
of these in the ASAS-3 and for 2 northern/equatorial objects in the NSVS observations. In
two cases (HR~4049 and HD~93662), the ASAS-3 $V$-magnitudes showed large scatter that was
probably caused by saturation. Nevertheless, one characteristic minimum was still
identifiable for HR~4049 (see later). In addition, we also downloaded ASAS-3 data for 7
other pulsating stars that do have periods listed in Table\ A.2. of De Ruyter et al.
(2006), so that we could check the consistency of our results with those in the
literature.

\begin{table}
\begin{centering}
\caption{\label{sample} Stars in the sample and properties of the photometric datasets}
\begin{scriptsize}
\begin{tabular}{lllrll}
\hline
IRAS& Other name & Sp. & $\langle V \rangle$ & JD from...to & $N_{\rm obs}$\\
    &           &  type   &         & 2450000+           &           \\
\hline
05208$-$2035 &          &M0e+F & 9.46 & 1868...3869 & 336\\
06072$+$0953 & CT~Ori & F9I & 10.47 & 2384...3865 & 168\\
06160$-$1701 & UY~CMa & G0V & 11.03 & 1868...3881 & 332\\
06176$-$1036 & HD~44179 & F1I & 8.84 & 1868...3880 & 285\\
06472$-$3713 & ST~Pup & F7I & 10.01 & 1868...3896 & 346\\
07140$-$2321 & SAO~173329 & F5I & 10.64 & 1868...3898 & 394\\
07284$-$0940 & U~Mon  & GoI & 6.46 & 1868...3896 & 299\\
08011$-$3627 & AR~Pup & F0I & 9.38 & 1878...3897 & 327\\
08544$-$4431 &          & F3 & 9.17 & 1868...3896 & 377\\
09060$-$2807 & BZ~Pyx   & F5 & 10.86 & 1868...3898 & 328\\
09144$-$4933 &          & G0 & 13.78 & 1868...3896 & 240\\
09256$-$6324 & IW~Car & F7I & 8.50 & 1868...3897 & 1425\\
09400$-$4733 &          & M0 & 10.56 & 1868...3896 & 327\\
09538$-$7622 &          & G0 & 11.93 & 1870...3898 & 356\\
10158$-$2844 & HR~4049 & A6I & 6.00 & 1871...3182 & 95\\ 
10174$-$5704 &         & K:rr & 11.30 & 1868...3190 & 270\\
10456$-$5712 & HD~93662 & K5 & 6.60 & 1868...3223 & 421\\
11000$-$6153 & HD~95767 & F0I & 8.79 & 1869...3231 & 218\\
11472$-$0800 &          & F5I & 11.58 & 1871...3189 & 167\\
             &          &     &       & 1274...1634 & 114$^*$\\          
12222$-$4652 & HD~108015 & F3Ib & 7.95 & 1954...3239 & 137\\
15469$-$5311 &           & F3 & 10.55 & 1936...3262 & 257\\
17038$-$4815 &           & G2p(R)e & 10.53 & 1933...3288 & 251\\
17233$-$4330 &           & G0p(R) & 12.26 & 1936...3260 & 178\\
17243$-$4348 & LR~Sco    & G2 & 10.50 & 1933...3260 & 215\\
17530$-$3348 & AI~Sco & G4I & 9.83 & 2417...3293 & 118\\
18123$+$0511 &           & G5 & 10.14 & 2159...3292 & 121\\
19125$+$0343 &           & F2 & 10.16 & 2138...3896 & 209\\
19157$-$0247 &           & F3 & 10.72 & 1979...3896 & 235\\
20056$+$1834 & QY~Sge    & G0 De & 10.8$^*$ & 1277...1632 & 120$^*$\\ 
22327$-$1731 & HD~213985 & A2I & 8.92 & 1871...3897 & 284\\
\hline
\end{tabular}
\vskip3mm
$^*$ -- unfiltered magnitudes from the NSVS project
\end{scriptsize}
\end{centering}
\end{table}

The main properties of the final sample of 30 stars are summarized in Table\ \ref{sample}.
The mean time-span of the ASAS-3 light curves is 1682 days, although for 14 stars the
time-base extends over 2000 days. With 300 datapoints in an average light curve, there is
one ASAS-3 observation every few days, giving excellent coverage for stars with periods
between 10 and 200 days. One of the two NSVS objects, QY~Sge, was observed 120 times over
the 1 year of operation, giving only a meagre picture of the variability, while NSVS data
for IRAS~11472$-$0800 were useful in supplementing the information about the period change
of the star.

\begin{figure*}
\begin{center}  
\leavevmode
\includegraphics[width=16cm]{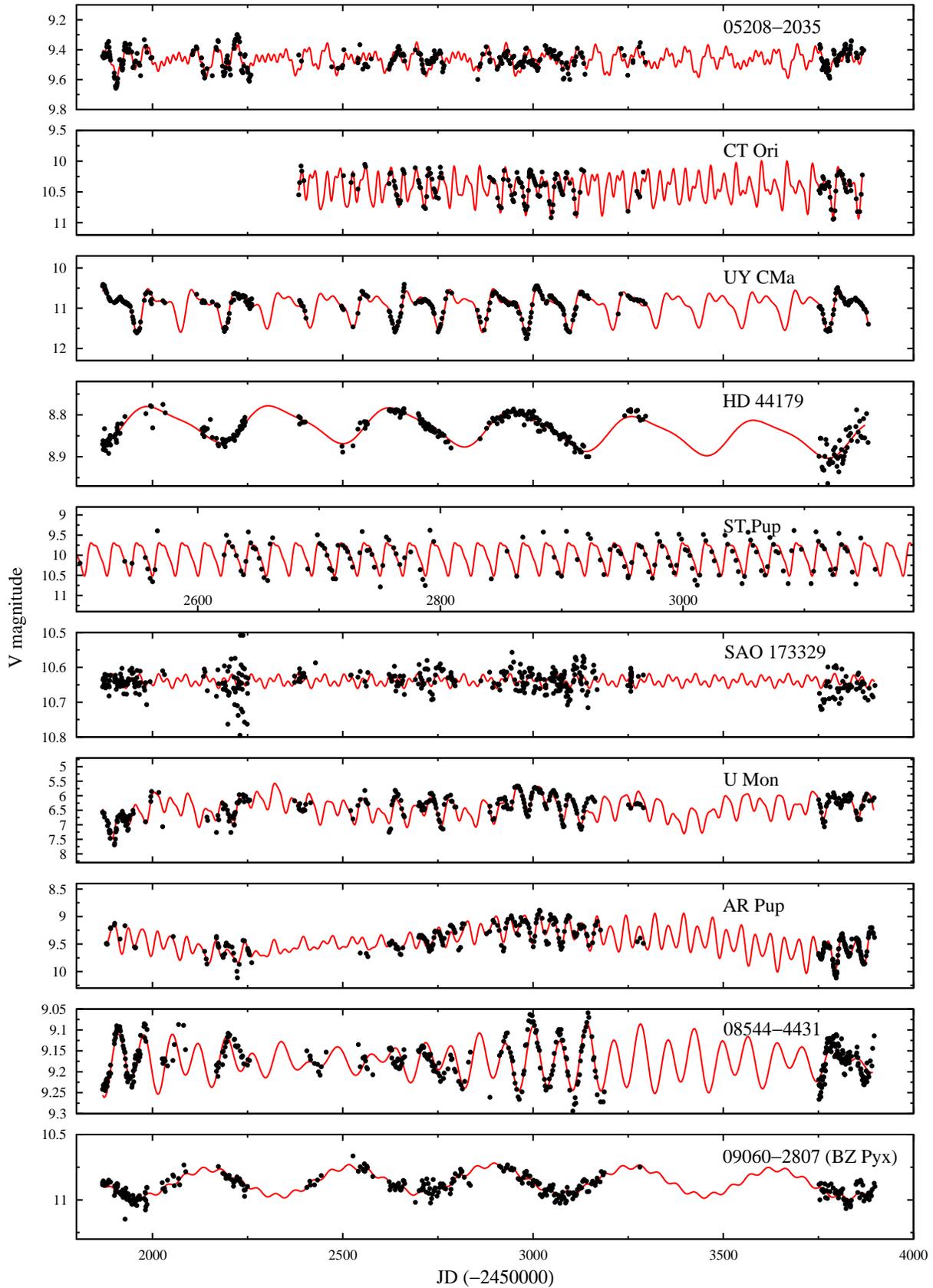}
\end{center}
\caption{ASAS $V$-band light curves of post-AGB stars. Solid lines denote low-order harmonic 
fits. For clarity we expanded the time-axis for ST~Pup, thus showing only half of
the ASAS data.}
\label{lc1}
\end{figure*}

\begin{figure*}
\begin{center}  
\leavevmode
\includegraphics[width=16cm]{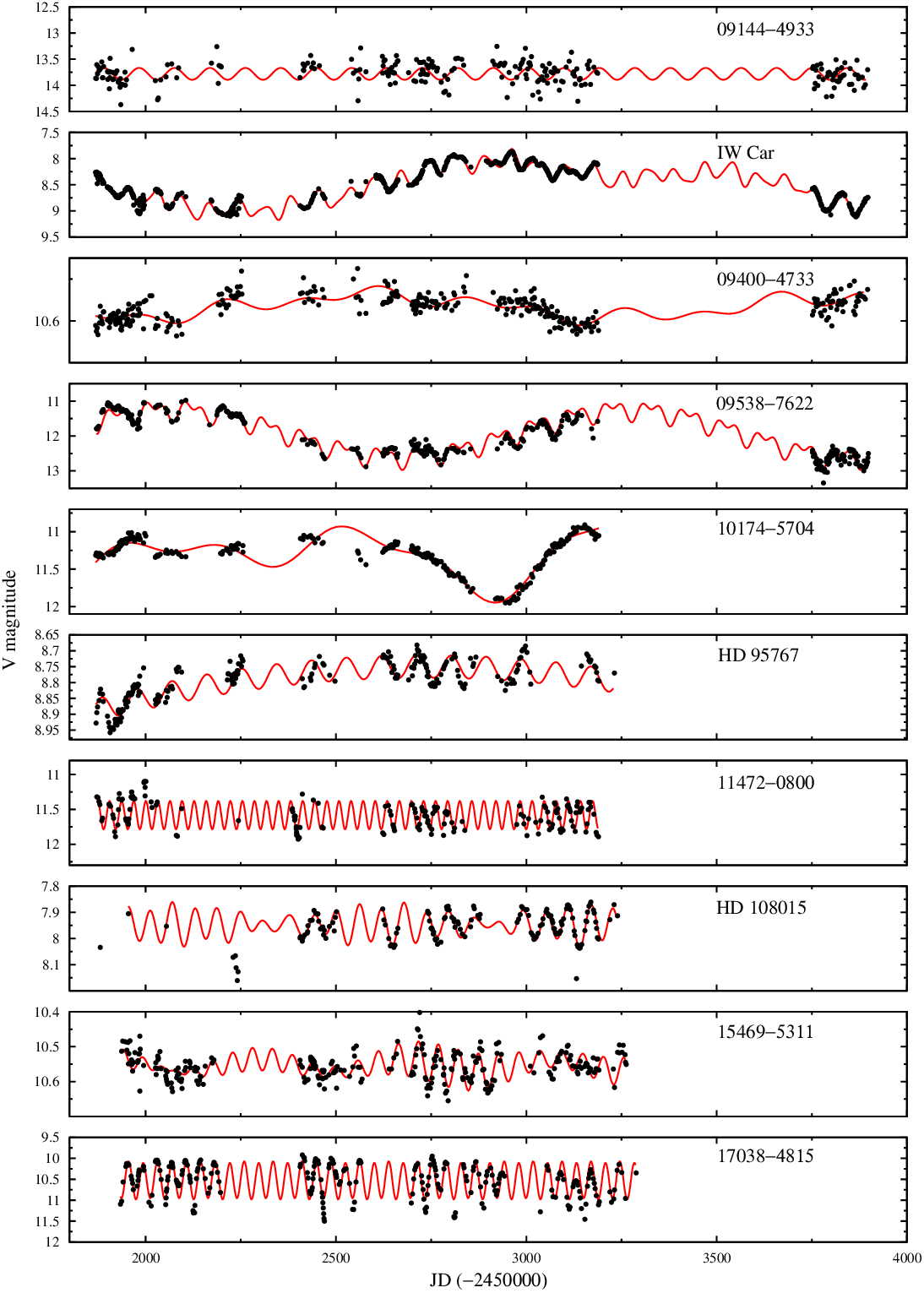}
\end{center}
\caption{ASAS $V$-band light curves and harmonic fits of post-AGB stars.}
\label{lc2}
\end{figure*}

\begin{figure*}
\begin{center}  
\leavevmode
\includegraphics[width=16cm]{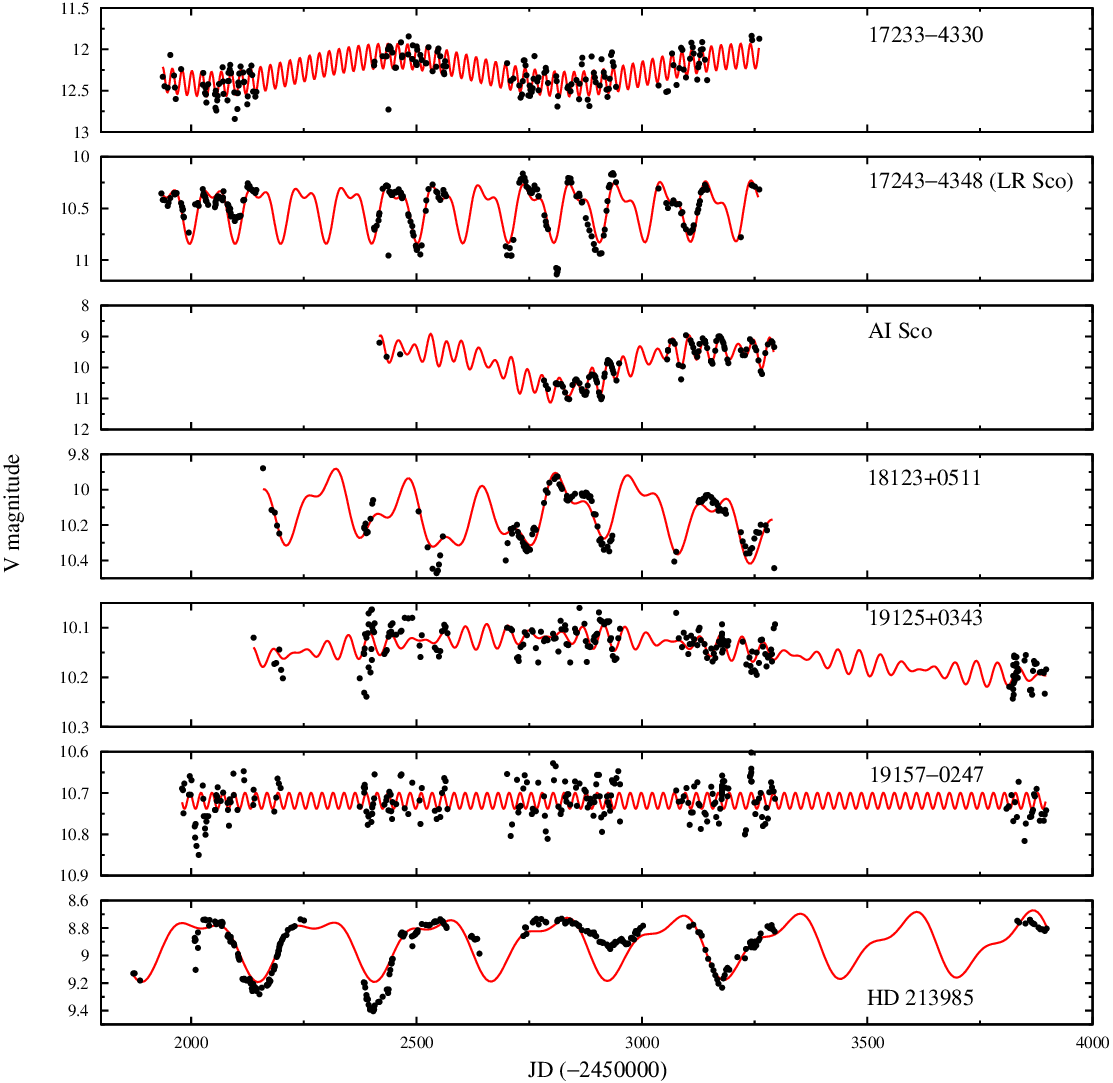}
\end{center}
\caption{ASAS $V$-band light curves and harmonic fits of post-AGB stars.}
\label{lc3}
\end{figure*}

We determined periods by iterative sine-wave fitting using {\sc Period04} of Lenz \& Breger
(2005). In several stars the variations are too complex for simple stationary Fourier-fits. In
those cases we fitted harmonic components with variable frequencies using our own codes.
Examples of power spectra are plotted in Fig.\ \ref{power}.
The uncertainties of the determined periods were conservatively estimated
from the spectral window by taking its full width at half maximum as the error 
in frequency.
For the RV~Tauri type stars,
after finding the mean period, we fitted a fixed set of integer multiplets of the mean
frequency to describe the highly non-sinusoidal light curve shapes. That was an approximation
for most RV~Tau stars, because their light curve cycles are well-known for not being strictly
repetitive. In Figs.\ \ref{lc1}--\ref{lc3} we show 27 of the resulting  light curve fits. 

\begin{table*}
\begin{centering}
\caption{Stellar parameters (taken from De Ruyter et al. 2006), $V$-band amplitude (to the nearest 0.05-0.1 mag) 
and the determined periods with 
$S/N>4$. Key to the variability types: `RVa' -- RV~Tauri pulsator with constant mean brightness;
`RVb' -- RV~Tauri pulsator with varying mean brightness; Cep II -- Population II Cepheid; 
`pulsating' -- low amplitude semiregular
or multiply periodic variable; `binary' -- photometric variations with the orbital period;
Irr -- variable without well-defined pattern (irregular).}
\begin{tabular}{llllrrlllr}
\hline
IRAS& Other name & $T_{\rm eff}$ & $\log~g$ & [Fe/H] & $\Delta V$ & Period(s)      & Type of      & Period(s)    & ref.\\
    &            &                &          &        &  [mag]     & (this study) [d] & variability & (literature) [d]  &        \\
\hline
05208$-$2035 &            & 4000 & 0.5 &  0.0 & 0.2 & 58.8$\pm$0.9, 47.6$\pm$0.6 & pulsating & --- & -- \\
            &             &      &     &      &     & 117$\pm$2.5, 15.6$\pm$0.7  &   & &\\
06072$-$0953 & CT~Ori     & 5500 & 1.0 & $-$2.0 & 0.8 & 67.3$\pm$1 & RVa& 135.52 & 1 \\
06160$-$1701 & UY~CMa     & 5500 & 1.0 & 0.0   & 1.1 & 113.7$\pm$2.5 & RVa/Cep II & 113.9 & 1 \\
06176$-$1036 & HD~44179   & 7500 & 0.8 & $-$3.3 & 0.1 & 319$\pm$20 + 2nd harmonic & binary & 318$\pm$3 & 2\\
06472$-$3713 & ST~Pup     & 5750 & 0.5 & $-$1.5 & 1.1 &  changing from  18.5 to 19.2 d & Cep II & 18.88 & 1\\
07140$-$2321 & SAO~173329 & 7000 & 1.5 & $-$0.8 & 0.05 & $\sim$60, $\sim$24 & Irr & --- & -- \\
07284$-$0940 & U~Mon      & 5000 & 0.0 & $-$0.8 & 1.1 & 91.3$\pm$2.3 & RVb & 92.26 & 1 \\
08011$-$3627 & AR~Pup     & 6000 & 1.5 & $-$1.0 & 0.5 & 76.4$\pm$4, 1250$\pm$300 & RVb & 75 & 1\\
08544$-$4431 &            & 7250 & 1.5 & $-$0.5 & 0.12 & 72.3$\pm$1.3, 68.9$\pm$1.2, 133$\pm$1 & pulsating & 499$\pm$3, 72 or 90 & 3\\             
09060$-$2807 & BZ~Pyx     & 6500 & 1.5 & $-$0.5 & 0.2 & 372$\pm$35, 48$\pm$6 & RVb & --- & -- \\           
09144$-$4933 &            & 5750 & 0.5 & $-$0.5 & 0.6 & 93$\pm$2 & pulsating & --- & -- \\
09256$-$6324 & IW~Car     & 6700 & 2.0 & $-$1.0 & 0.4 & 72.1$\pm$1, 1400$\pm$500, 42$\pm$0.5 & RVb & 67.5 & 1\\        
09400$-$4733 &            &      &     &        & 0.1 & $\sim$1300, $\sim$200 & pulsating & --- & -- \\              
09538$-$7622 &            & 5500 & 1.0 & $-$0.5 & 1.0 & 1190$\pm$300, 100.6$\pm$2.5 & RVb & --- & --\\            
10158$-$2844 & HR~4049   & 7500  & 1.0 & $-$4.5 & $>$0.5 & one characteristic minimum & binary & 434 & 4 \\ 
10174$-$5704 &           &       &     &     &  1.0& time-scale of 500--600 d & binary? & --- & -- \\           
10456$-$5712 & HD~93662  & 4250 & 0.5  & 0.0 & $>$0.5 &  two maxima $\sim$260 d apart, scatter & binary? & --- & -- \\
11000$-$6153 & HD~95767  & 7600 & 2.0 & 0.1  & 0.1 &  92$\pm$2 ($\sim$2000) & pulsating & --- & -- \\ 
11472$-$0800 &           & 5750 & 1.0 & $-$2.5 & 0.6 &  31.5$\pm$0.6 & Cep II &  --- & --\\           
12222$-$4652 & HD~108015 & 7000 & 1.5 & $-$0.1 & 0.1 & 61$\pm$1.9, 55$\pm$1.6 & pulsating & --- & --\\  
15469$-$5311 &           & 7500 & 1.5 & 0.0 & 0.1 & 54.4$\pm$1, 384$\pm$50, 49.1$\pm$0.8 & pulsating & --- & -- \\           
17038$-$4815 &           & 4750 & 0.5 & $-$1.5 & 1.5 &75.9$\pm$1.9 & RVa & --- & -- \\           
17233$-$4330 &           & 6250 & 1.5 & $-$1.0 & 0.5 & 735$\pm$230, 44$\pm$0.8 & RVb & --- & -- \\          
17243$-$4348 & LR~Sco    & 6250 & 0.5 & 0.0 & 0.8 & 100.5$\pm$4 & RVa & 104.4 & 5 \\ 
17530$-$3348 & AI~Sco    & 5000 & 0.0 & 0.0 & 1.0 & 71.6$\pm$5.1, $\approx$ 870 & RVb & 71.0 & 1 \\         
18123$+$0511 &           & 5000 & 0.5 & 0.0 & 0.6 & 172$\pm$15 & RVa & --- & -- \\
19125$+$0343 &           & 7750 & 1.0 & $-$0.5 & 0.05 & $\sim$2300, time-scale of 50 d & pulsating & --- & --\\
19157$-$0247 &           & 7750 & 1.0 & 0.0 & 0.05 & $\sim$25 d & Irr &  --- & -- \\          
20056$+$1834 & QY~Sge    & 5850 & 0.7 & $-$0.4 & -- & four minima at $\sim$60 d separation & pulsating & $\sim$50 & 6 \\ 
22327$-$1731 & HD~213985 & 8250 & 1.5 & $-$1.0 & 0.4 & 261.5$\pm$17 + 2nd harmonic & binary & 254 & 7 \\ 
\hline
\end{tabular}
\end{centering}
\vskip3mm
1 -- De Ruyter et al. (2006); 2 -- Waelkens et al. (1996); 3 -- Maas et al. (2003); 4 -- Waelkens et al. (1991);
5 -- Kholopov et al. (1985-1988); 6 -- Menzies \& Whitelock (1988); 7 -- Whitelock et al. (1989)
\label{periods}
\end{table*}

\section{Results}

All of the objects in the sample turned out to be photometrically variable (note that  the
ASAS-3 catalog contains all the observed stars, not only the variable ones). It is apparent
from Figs.\ \ref{lc1}--\ref{lc3} that the sample has a very rich photometric behaviour. In
most cases we could determine at least one period, while in 13 stars we infer multiple
periodicity with a wide range of period ratios. In cooler (and hence larger amplitude) stars,
the long secondary periods correspond to the so-called RVb phenomenon, i.e. long-term changes
of the mean brightness, coupled with the short-term RV~Tauri pulsations. We identify five new
genuine RV~Tauri variables, whose RV~Tauri nature was suggested by their position in the IRAS
two-colour diagrams (Lloyd-Evans 1997). In a few objects we see several cases of beating,
caused by multiple frequencies with ratios between 1.0 and 2.0. In those cases we determined 2
or 3 periods, although we note that the time-span of the data is not long enough to make a
distinction between non-repetitive quasi-regularity  and genuine multiple periodicity. 

We present stellar parameters and periods in Table\ \ref{periods}. Detailed notes on 
each star are given in Appendix\ A.  

\section{Discussion}

Our study aimed at determining periods for a well-defined sample of post-AGB stars,
which includes confirmed and candidate binary objects. The homogeneous set of ASAS-3
$V$-band data allowed us to study photometric variability with an extended coverage of
this interesting evolutionary phase. The immediate result is an improved knowledge
about pulsations and orbital brightness modulations in a relatively large set of
post-AGB stars. Let us recall that the most extensive catalogue contains about 220
objects (Szczerba et al. 2001), so that our 30 stars form a significant fraction of
post-AGB stars known to date. By restricting our investigation to the De Ruyter et al.
(2006) sample, we ensured that the properties of all stars are known to similar
accuracy. 

Almost half of the sample comprises RV~Tau-type pulsating variables, of which five are
new identifications. Assuming that the RVb phenomenon is a signature of orbital
modulations, we have measured the orbital periods for three stars (BZ~Pyx: 372 d; 
IRAS~09538$-$7622: 1190 d; IRAS~17233$-$4330: 735 d). Their confirmation will require
long-term spectroscopic monitoring. In three more RVb stars (AR~Pup, IW~Car and AI~Sco)
we measured modulation periods that are consistent with those in
the literature.

We find two new candidates for binary modulations (IRAS~10174$-$5704 and HD~93662) to
add to the well-known  examples of HD~44179, HR~4049 and HD~213985. Unfortunately, in
neither case are the data long or accurate enough to derive orbital periods. We can only
say that the time-scales are a few hundred days. Of the known binary variables,
HD~213985 is most interesting for future monitoring because of the cycle-to-cycle
changes of the light curve shape. This behaviour suggests on-going changes in the
circumstellar shell, and high-resolution observations may shed new light on the
mass-loss processes in post-AGB binaries.

\subsection{Stars with period variations}

One star in the sample, the peculiar Pop. II Cepheid ST~Pup, has long been 
known to show long-term fluctuations in period that are not consistent with
evolutionary theories (Wallerstein 2002) and for which there is no firm theoretical 
explanation. ST~Pup might be similar to V725~Sgr, which first increased its period from
12~d to 21~d between 1926 and 1935 and then, most recently, up to 90~d, representing a
blue loop in the HRD from the AGB to the instability strip and back again (Percy et al.
2006).

\begin{figure}
\begin{center}  
\leavevmode
\includegraphics[width=8cm]{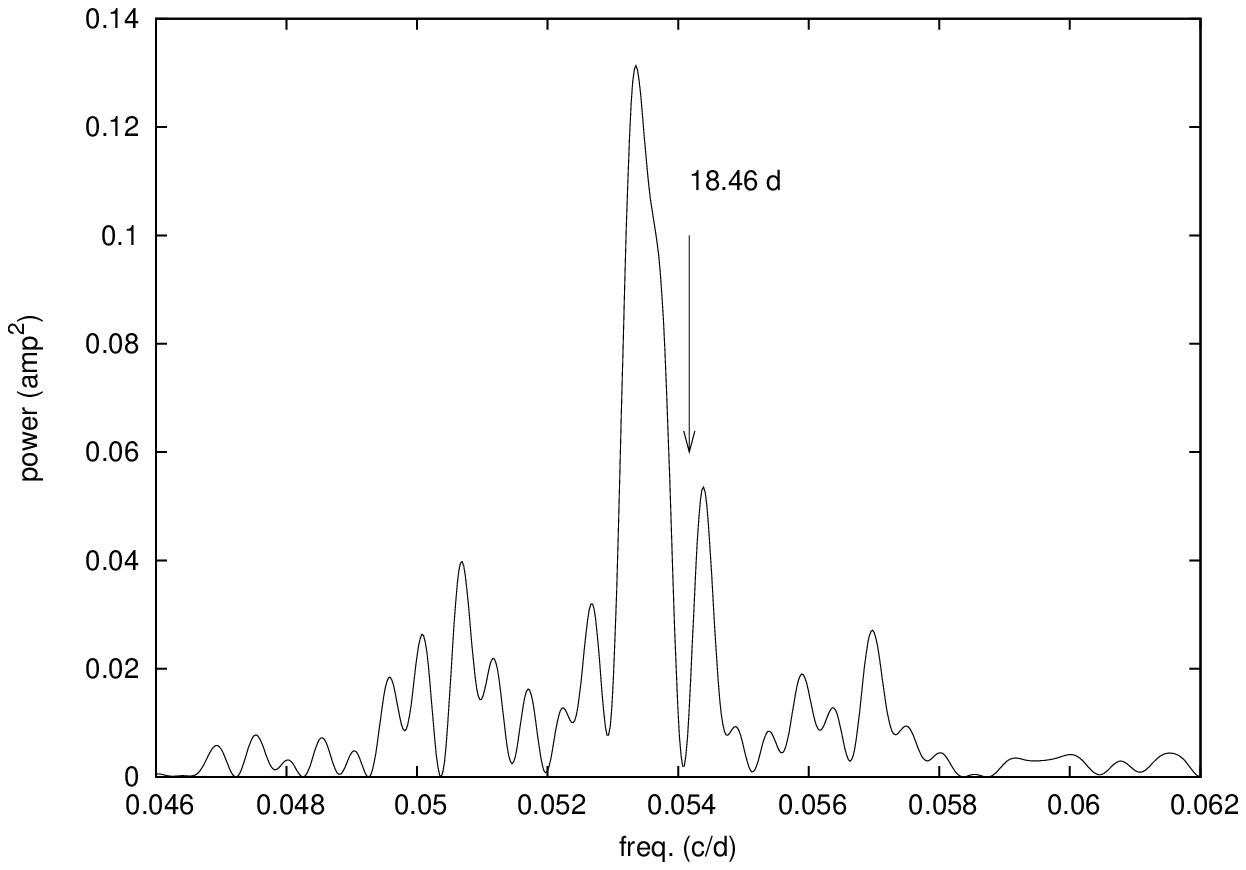}
\includegraphics[width=8cm]{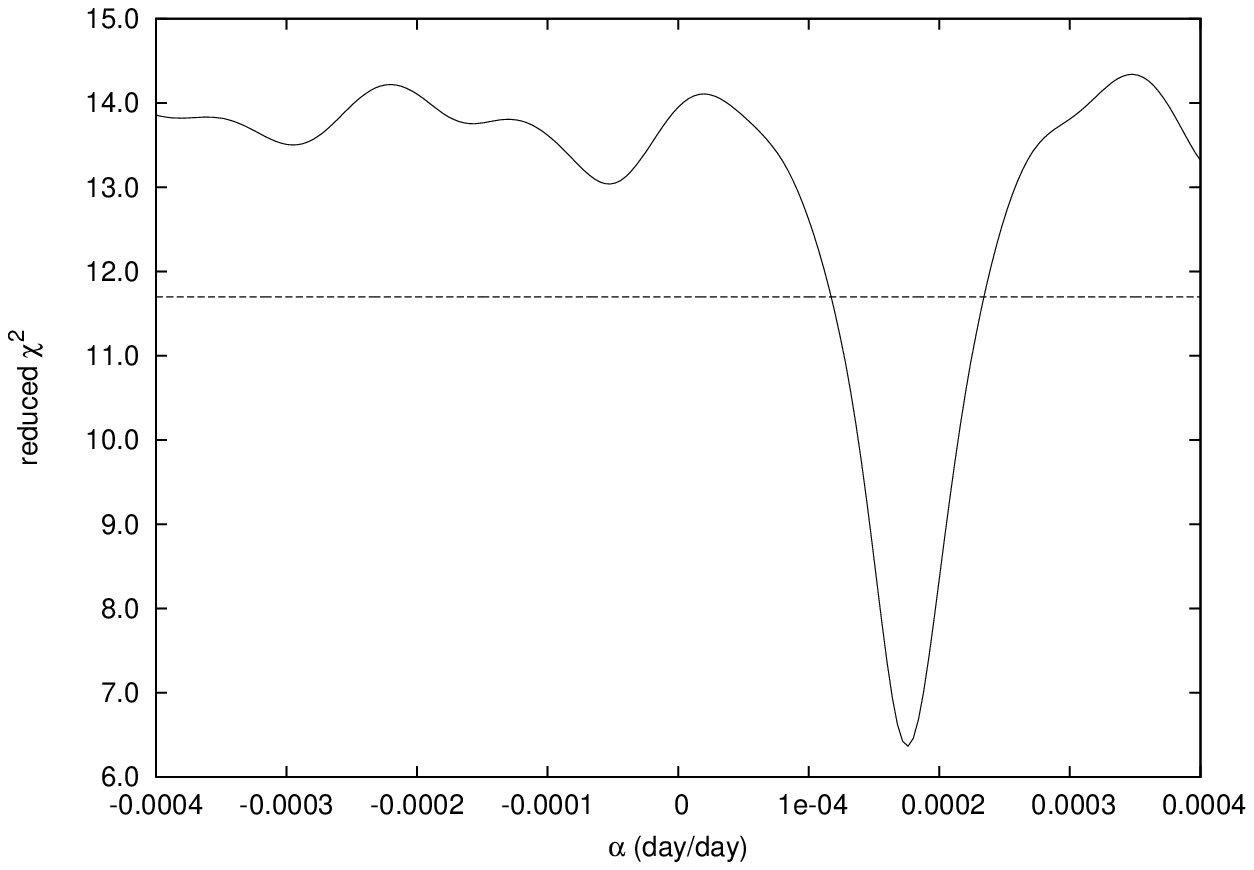}
\end{center}
\caption{{\it Top panel:} a close-up view of the power spectrum of ST~Pup. The arrow marks 
the mean period around the early 1990s (Gonzalez \& Wallerstein 1996). Side-lobe peaks
come from the spectral window. {\it Bottom panel:} reduced $\chi^2$ as a function of the 
linear period evolution parameter $\alpha$, assuming $P(t) \sim P_0 + \alpha t$. 
The dashed line shows the $\chi^2$ of the best monoperiodic fit.}
\label{stpup}
\end{figure}

The ASAS photometry indicates a strongly changing pulsation period
with a linear increase of $\alpha=dP/dt=1.75\times10^{-4}$ day/day 
(see Fig.\ \ref{stpup}). Note that this result is based on the
simple assumption of linear period change, which is not valid for the whole history of
ST~Pup. Gonzalez \& Wallerstein (1996) determined the orbital period of this system from
radial velocity data to be 410.4 days. The value of orbital periodicity, however,
strongly depends on the radial velocity variations caused by the pulsation. When
decomposing the radial velocity changes into pulsational and orbital components, it is
essential to use the correct pulsation period derived from photometry. From the present
period study it is obvious that the pulsation period of 18.4622 days used by Gonzalez \&
Wallerstein was incorrect for the epoch of their radial velocity observations. Instead,
the value of 18.4298 day is a reasonable approximation. Decomposing the radial velocity
data with this latter period, a much shorter period emerges for the orbital periodicity,
instead of 410.4 days, the value of 25.67 days fits much better the radial velocity data
prewhitened with the variations of pulsational origin. Such a short orbital period (if
confirmed) means that the secondary star can influence the pulsation significantly and
offers a natural explanation for the instability of the pulsation period of ST~Puppis.

\begin{figure}
\begin{center}  
\leavevmode
\includegraphics[width=8cm]{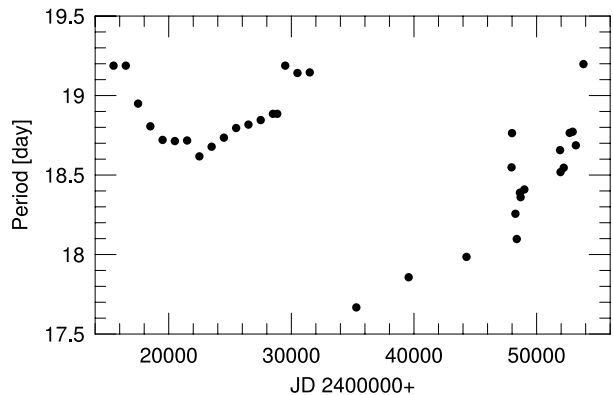}
\end{center}
\caption{The pulsation period of ST~Puppis varies erratically. This Type~II 
Cepheid shows both smooth and violent period changes incompatible with evolutionary 
effects.}
\label{stpupper}
\end{figure}

These strong period changes in ST~Puppis occurring on a short time scale prompted us to
study the behaviour of the pulsation period of this particular Cepheid using all
available photometric data. The period for the first half of the 20th century was
published by Payne Gaposchkin (1950). In general, the period was longer than now and the
changes in it were apparently less sudden as compared with the recent behaviour.
Reliability of these period values is supported by an independent period determination
made by Hoffmeister (1943) based on another series of photographic observations. From the
mid-20th century on, data have been obtained with photoelectric equipment and, more
recently with CCD cameras. Because the individual data are all accessible, a homogeneous
treatment is possible for determining the instantaneous pulsation period. The data series
analysed included (in the order of the epoch of observations) the observations by Irwin
(1961), Walraven, Muller \& Oosterhoff (1958), Landolt (1971), Harris (1980), Kilkenny
et~al. (1993), the Hipparcos satellite (ESA 1997), and Berdnikov \& Turner
(2001). At first, the periods were determined separately for each dataset (photometric
samples covering several years were divided into reasonably short segments). The temporal
dependence of the pulsation period is plotted in Figure\ \ref{stpupper}. The uncertainties
of the individual period values are smaller than the size of the circles. It is seen that
violent period changes are superimposed on the smooth (yet strong) variations of the
pulsation period. Most conspicuous jumps in the period occurred between JD\,2\,432\,000
and JD\,2\,435\,000,  near JD\,2\,449\,000, and between JD\,2\,453\,000 and
JD\,2\,454\,000. This erratic behaviour is incompatible with stellar evolution across the
instability strip. Whether it has to do with the presence of a companion is a question
that must be answered by dedicated observations.

Another Pop. II Cepheid, IRAS~11472$-$0800 exhibited quite strong phase 
modulations without a clear long-term trend (see the Appendix for details).
We find no star in the sample with FG~Sge-like period change that might be due to
the final helium flash during post-AGB evolution (e.g. Bl\"ocker \& Sch\"onberner 1997).
A comprehensive analysis of all post-AGB stars identified to date in the available
photometric surveys, most importantly in the ASAS database, will be an important future
task.       

\begin{figure}
\begin{center}  
\leavevmode
\includegraphics[width=8cm]{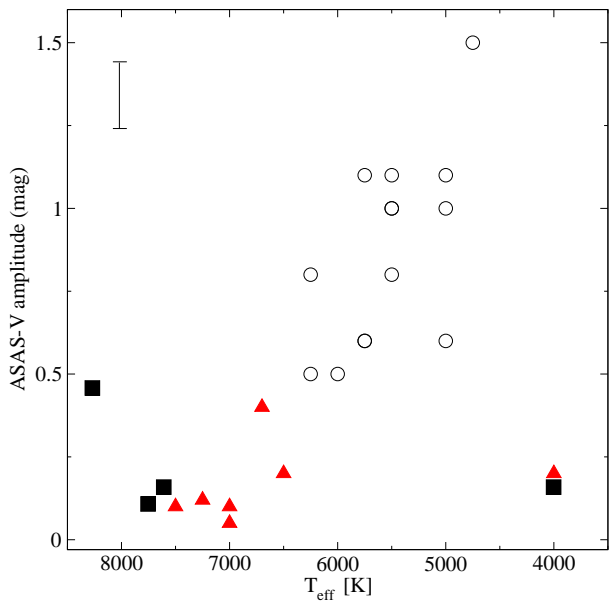}
\end{center}
\caption{$V$-band amplitude as a function of effective temperature for post-AGB
variables. Open circles: single periodic stars; triangles: multiperiodic/semiregular
stars; squares: variability due to orbital motion. The bar in the upper left corner 
shows the typical cycle-to-cycle amplitude variation in the pulsating stars.}
\label{teffamp}
\end{figure}

\subsection{The pulsations of post-AGB stars}

In Fig.\ \ref{teffamp} we show peak-to-peak amplitudes of the ASAS-3 light curves versus
effective temperature (cf. fig.\ 1 of Lloyd Evans 1999). For the RVb stars, we first
removed the long-term  modulation. Since the effective temperature is a good indicator
of the post-AGB evolutionary phase, the correlation between the $V$-band amplitudes and
$T_{\rm eff}$  is a spectacular indication of how post-AGB stars evolve through the
instability strip. The temperatures were taken from De Ruyter et al. (2006), who
determined a consistent set of stellar photospheric temperatures for the majority of the
sample from high-resolution spectra. For RV~Tauri stars they deduced the parameters from
spectra taken preferentially around the maximum light, when the effects of the molecular
bands are weaker. 
In Fig.\ \ref{teffamp} we made a distinction between singly periodic and
multiply periodic (or possibly semiregular) variables to show how pulsation behaviour
changes on the two edges of the instability strip. Note that for the coolest RV~Tau
stars the $V$-band amplitude is enhanced by the temperature sensitivity of the molecular
bands in the optical spectrum. However, for temperatures above 5000 K, where most of the
stars lie, we expect the effect to be small.

We draw several conclusions based on Fig.\ \ref{teffamp}. The position of the 
instability strip is very well-defined by the boundaries at $\approx$4500 K and  6500 K.
This is where the higher amplitude Type II Cepheid/RV~Tauri pulsations are excited by
the $\kappa$-mechanism. For stars that have left the instability strip, the pulsations
are much less dynamic and regular.  However, the lack of a sharp edge on the hot side is
exactly as predicted by model calculations (Aikawa 1993, Gautschy 1993). It is very
interesting that a similar behaviour may occur for the red edge, too, although the
evidence is based on a single object (IRAS~05208$-$2035). Despite the lack of stars
between 4000 K and 4750 K, the asymmetric amplitude distribution suggests a narrower red
edge for the instability strip, which might be understood in terms of a stronger damping
by convection in the cooler stars. 

\begin{figure}
\begin{center}  
\leavevmode
\includegraphics[width=8cm]{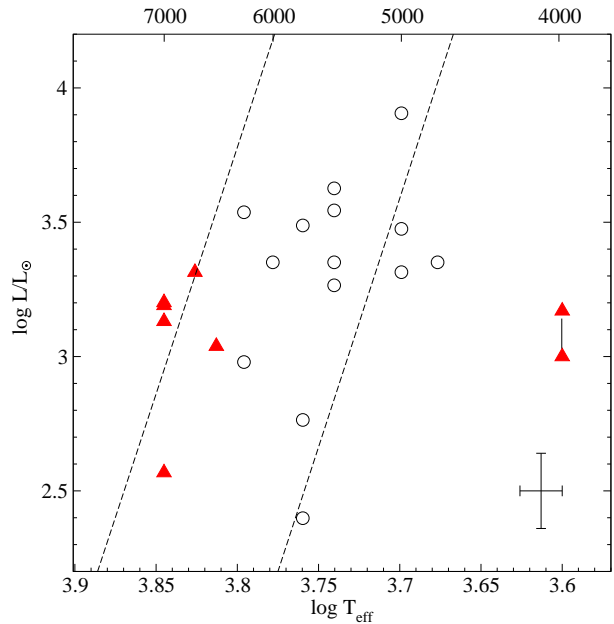}
\end{center}
\caption{The empirical HRD of the pulsating sample, using the RV~Tau period--luminosity
relation in the Large Magellanic Cloud. The dashed lines show the edges of the 
classical instability strip, taken from Christensen-Dalsgaard (2003). The error bars
in the lower right corner represent $\pm3\%$ error in effective temperature (about 200 K in
the range shown) and the $\pm0.35$ mag standard deviation of the LMC P--L relation.}
\label{hrd}
\end{figure}

To place our stars in the Hertzsprung--Russell diagram, we have used the
period--luminosity relation for RV~Tauri stars in the Large Magellanic Cloud derived by
Alcock et al. (1998), which has been used by De Ruyter et al. (2006) for estimating the
luminosities of the pulsating objects in their sample. We used the same P--L relation to
calculate $V$-band absolute magnitudes:

$M_V=2.54(\pm0.48)-(3.91\pm0.36) \log (P/2)$, \hskip3mm $\sigma=0.35$

\noindent for variables with $P/2>12.6$ days. To calculate luminosities, we took $BC_V$
values from Table\ A.2 in De Ruyter et al. (2006). The resulting empirical HRD is shown
in Fig.\ \ref{hrd}, where we also show the boundaries of the classical instability strip,
adopted from Christensen-Dalsgaard (2003). We did not include stars hotter than 7000 K
because the applicability of the RV~Tau P--L  relation becomes doubtful for stars that
have parameters outside the calibrating sample. This is also true for the coolest
multiply periodic star on the right-hand side in Fig.\ \ref{hrd} (IRAS~05208$-$2035), but
its position in the plot could still be interesting, so we applied the P--L relation to
both of the periods. 

Fig.\ \ref{hrd} reveals a very nice agreement with the classical instability strip. 
Using other models of the Cepheid instability strip (e.g. Saio \& Gautschy 1998; Baraffe
et al.  1998; Bono et al. 2005) gives slight changes in the slope and exact
locations of the theoretical edges but does not change the overall agreement. This
means  that the LMC RV~Tau P--L relation and the adopted spectroscopic temperature
estimates, both free of any theoretical assumption on the pulsations,  yield a
consistent physical picture of post-AGB stars evolving through the instability strip
(see also Lloyd Evans 1999). Interestingly, there is a suggestion for a more extended
red edge from the outside  positions of the three coolest and largest amplitude RV~Tau
stars (near the location $\log T_{\rm eff}=3.7$, $\log L/L_\odot=3.4$). This becomes
more evident if we consider that the temperatures of those stars were determined in the
hottest states, so that their mean temperatures are even cooler. We interpret this
subtle difference as one caused by the different stellar structure of Cepheids and
post-AGB stars. They may overlap in temperature and luminosity, but Cepheids are massive
stars (5--8 M$_\odot$), which means their density profile will be different from
that of the post-AGB objects. Therefore, we do not expect the same efficiency of
excitation at a given position of the HRD, especially at the edges of the instability
strip.  

Finally, some of the objects on the blue edge of the instability strip present evidence
for multiply periodic oscillations with period ratios around 1.1 (cf. Fig.\
\ref{power}). The simplest explanations are the presence of radial and non-radial modes
or of very high-order radial modes. The former case could be interesting in the search
for mechanisms that may lead to non-spherical mass-loss (e.g. Soker 2000). These are
also the objects for which asteroseismology could be very rewarding, provided that
realistic models are calculated for a wide range of parameters (see Fokin et al.
2001).  

\section*{Acknowledgments} 

This work has been supported by a University of Sydney Postdoctoral Research Fellowship,
the Australian Research Council, the Hungarian OTKA Grants \#T042509, \#T046207 and the
Ma\-gyary Zolt\'an Public Foundation for Higher Education.  The ASAS and NSVS online
catalogues and the NASA ADS Abstract Service were used to access data and references.
This research has made use of the SIMBAD database, operated at CDS-Strasbourg, France.

\appendix

\section{Notes on individual stars}

Except where noted, nothing has previously been published on the variability of most of
these stars.

\paragraph*{IRAS 05208$-$2035:} This star belongs to the candidate RV~Tauri stars based on the
position in the IRAS colour-colour diagram of Lloyd Evans (1997). Previously, Hashimoto (1994)
listed the star among oxygen-rich AGB stars. The light curve (Fig.\ \ref{lc1}) is dominated by
short-period, complex variations with time-scales from 16~d to 120~d, which are hardly
compatible with typical AGB stars. None of these periods comes from a coherent signal that
might have indicated orbital motion.

\paragraph*{CT~Ori:} De Ruyter et al. (2006) listed a period of 135 d, which can be
traced back to Hoffmeister (1930), but which is not correct. It was already pointed out
by Horowitz (1986) that the period is more likely to be half or even a quarter of
that. The ASAS data (Fig.\ \ref{lc1}) clearly shows alternating
light curve with 67 days between two consecutive primary minima. 

\paragraph*{UY~CMa:} We confirm the catalogued period of $\sim$114 d.

\paragraph*{HD~44179:} This binary is the central star of the Red Rectangle nebula, with
an orbital period of 318$\pm$3 d. Waelkens et al. (1996) measured an asymmetric light
curve with 0.12 mag amplitude in $V$ and no colour variation over the optical range.
Their data, taken over six orbital cycles, revealed only minor light curve shape
changes, suggesting a stable configuration of the inner circumstellar disk. Waelkens et
al. (1996) proposed that photometric variability is not caused by variable extinction
due to the orbital motion (as was commonly assumed) but by the variation around the
orbit of the scattering angle of the light that is observed. Our analysis confirms the
period and, apart from a tiny zero-point shift of the photometry, we find a very similar
light curve to that measured by Waelkens et al. (1996) more than a decade earlier (Fig.\
\ref{lc1}). This
suggests stable conditions around the binary over the whole baseline of about two
decades. The power spectrum does not contain any excess power that could be indicative
of low-amplitude pulsations.

\paragraph*{SAO~173329:} We find that this star is variable at the 0.1 mag level. The
light curve (Fig.\ \ref{lc1}) is rather irregular; the listed ``periods'' are
time-scales of dimmings rather than real periods.

\paragraph*{U~Mon:} This is a well-known RV~Tau variable of the `b' subtype. The ASAS data
(Fig.\ \ref{lc1}) are not long enough to detect the $\sim$2600 d modulation of the mean
brightness, which was identified as the orbital period of the system (Pollard \& Cottrell
1995).  

\paragraph*{AR~Pup:} We confirm both the known 75 d pulsation period and 
the RVb modulation period of $\sim$1200 d (Raveendran 1999; see Fig.\ \ref{lc1}). 

\paragraph*{IRAS 08544$-$4431:} Maas et al. (2003) measured an orbital period of
499$\pm$3~d and found two periods of nearly equal probability, 71.9~d and 89.9~d. The
ASAS light curve (Fig.\ \ref{lc1}) confirms the shorter pulsation period (72~d) as well
as a slightly shorter one (69~d), producing a clear beat. The presence of a third
significant period ($S/N>4$ for 133~d) that is close to the twice the shortest one,
supports the multi-mode interpretation of the apparent amplitude modulation. Despite
covering 4 orbital cycles, the light curve does not show evidence for orbital brightness
variations. 

\paragraph*{BZ~Pyx:} The variable star designation and class SR: were given by
Kazarovets, Samus \& Durlevich (2001), who referred to a VSNET report by K. Takamizawa
in 2000, in which the discovery of new field variable stars was announced with no period
determination. The ASAS light curve (Fig.\ \ref{lc1}) suggests two periods
and resembles an
RVb type pulsation with long-term modulation, except that the pulsation amplitude and
period are far smaller than in typical RVb stars. The regularity of the modulation
suggests that the 372$\pm$1 d period corresponds to orbital motion.  

\paragraph*{IRAS 09144$-$4933:} The SIMBAD database lists this as an RV~Tau variable,
but no reference or period are given. The star is quite faint for the ASAS instrument,
which is the reason for the high point-to-point noise in the light curve (Fig.\
\ref{lc2}). Although the determined period would be typical for an RV~Tau star, the
light curve is too noisy for a firm classification. 

\paragraph*{IW~Car:} This star is a C-rich RVb variable with a 67.5--80 d pulsation and 1500 d
modulation period (Giridhar, Rao \& Lambert 1994). The latter is confirmed by the ASAS data,
while the short-period oscillations are best described with periods of  72~d and 42~d (Fig.\
\ref{lc2}). However, the shorter one, being close to the harmonic of the main period, might
only be caused by the non-sinusoidal shape of a non-stationary signal.  

\paragraph*{IRAS 09400$-$4733:} The light curve (Fig.\ \ref{lc2}) shows  slow
fluctuations but the overall behaviour does not seem to be periodic.

\paragraph*{IRAS 09538$-$7622:} We find this star to be a typical RVb type pulsating
star with well-defined modulation period close to 1200~d (Fig.\ \ref{lc2}).

\paragraph*{HR~4049:} Waelkens et al. (1991) found a periodicity of 434 d, which was
interpreted as an orbital period. The star is bright ($V<6$ mag) and the ASAS data seem
to have been affected by saturation. Nevertheless, one characteristic minimum was
observed around JD 2452800, which extended over 200 d and had a very similar shape to
those observed by Waelkens et al. (1991). 

\paragraph*{IRAS 10174$-$5704:} The ASAS light curve (Fig.\ \ref{lc2}) shows variations
up to 1 mag with a characteristic time-scale of 500-600 d. 

\paragraph*{HD~93662:} Lake (1963) suspected its variability (hence the designation
NSV~4975) but nothing specific has been published since then. The ASAS data have large
scatter, presumably from saturation effects, but two minima about 260 d apart are still
recognizable. 

\paragraph*{HD~95767:} The light curve (Fig.\ \ref{lc2}) clearly shows a small-amplitude oscillation with 92 d period and a
gradual change of the mean brightness that could indicate a modulation period of
$\sim$2000 d.  

\begin{figure}
\begin{center}  
\leavevmode
\includegraphics[bb=293 51 553 551,width=4.3cm,angle=270]{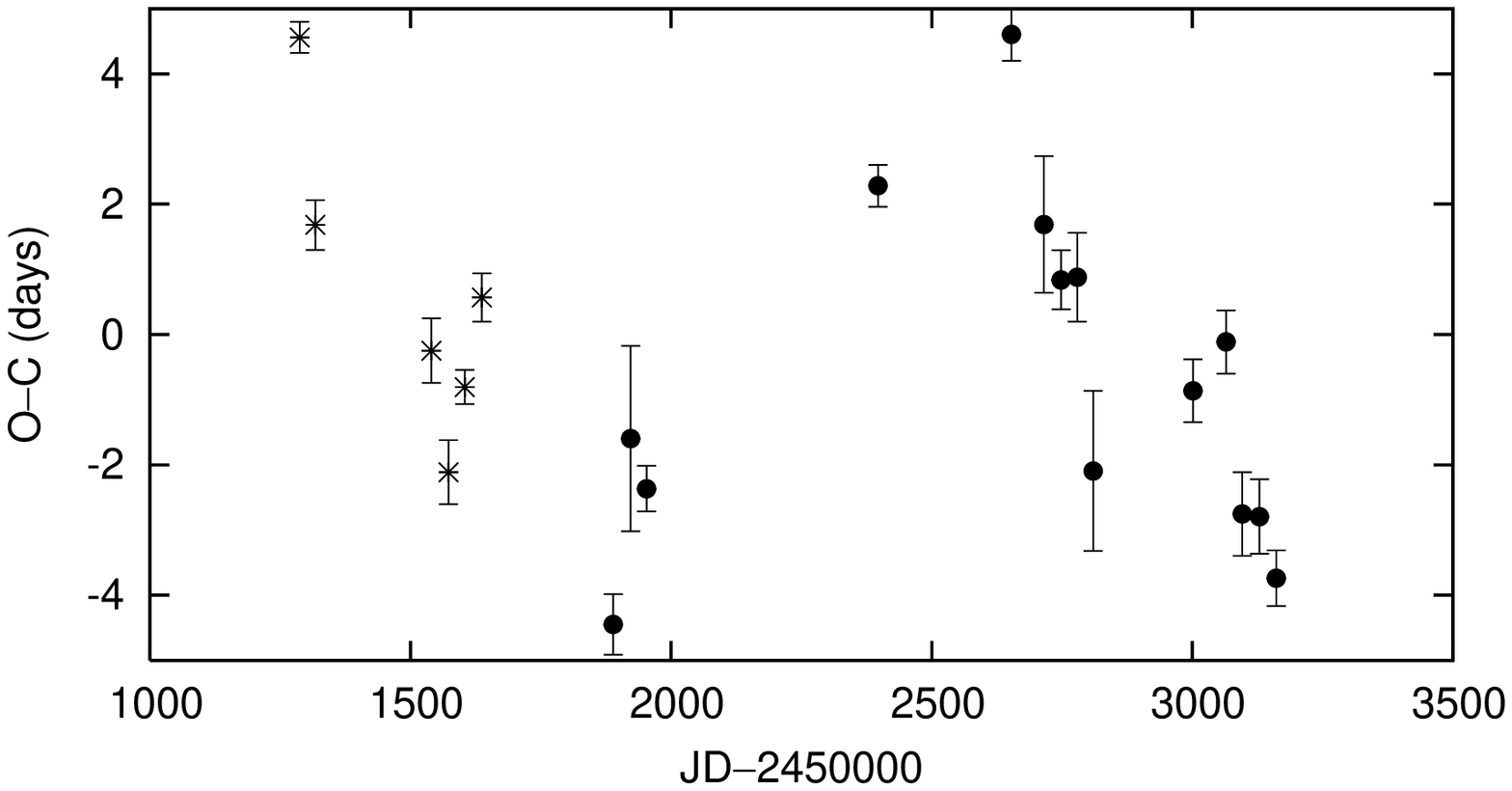}
\end{center}
\caption{Cycle-to-cycle phase variations of IRAS~11472$-$0800 from
NSVS (stars) and ASAS--3 (dots) data.}
\label{11472}
\end{figure}

\paragraph*{IRAS 11472$-$0800:} This is a relatively low-amplitude Type II Cepheid  with
31.5 d pulsation period (Fig.\ \ref{lc2}). We find quite strong
phase variations up to $\pm$5 d or 0.16 pulsation phase (Fig.\ \ref{11472}).

\paragraph*{HD~108015:} This is a bright, low-amplitude, multiply periodic pulsator
(Fig.\ \ref{lc2}) with no previous variability study.

\paragraph*{IRAS 15469$-$5311:} Very similar to HD~108015 in its variability (Fig.\
\ref{lc2}).

\paragraph*{IRAS 17038$-$4815:} This star has been included in the abundance analysis of
an extended sample of field RV~Tau stars (Giridhar et al. 2005), with no specific
variability information given. The light curve shows typical RV~Tau variations with 76 d
period (Fig.\ \ref{lc2}).

\paragraph*{IRAS 17233$-$4330:} We derive variability parameters for the first time.
Besides a long-term modulation (Fig.\ \ref{lc3}), the light
curve shows 0.5 mag oscillations with a formal period of 22 d. However, tiny differences
in minima suggest a doubled period of 44 d, which puts the star among the RVb type
variables.

\paragraph*{LR~Sco:} First misclassified as R~CrB star then realized it was a yellow
semiregular variable (Giridhar, Rao \& Lambert 1992). We determine a period that is close
to the one in the GCVS. The light curve shape is typical of an RV~Tau star (Fig.\
\ref{lc3}).

\paragraph*{AI~Sco:} We measure a very similar pulsation period than listed by De
Ruyter et al. (2006), which originates from Pollard et al. (1996). These authors measured
a long (RVb) period of 940 d, which is not entirely inconsistent with the ASAS data,
especially if we consider their short time-span ($<$900 d, see Fig.\ \ref{lc3}). 

\paragraph*{IRAS 18123$+$0511:} We find this to be an RV~Tau-like variable (Fig.\
\ref{lc3}). Its period is somewhat long and the amplitude is low for a typical RVa-type
star. 

\paragraph*{IRAS 19125$+$0343:} The SIMBAD database lists as a variable star of RV~Tau
type, but neither the period, nor the source of information is given. Besides a short
period and low amplitude fluctuation we also find a long-term modulation (Fig.\
\ref{lc3}).  

\paragraph*{IRAS 19157$-$0247:} We find tiny fluctuations just above the noise (Fig.\
\ref{lc3})  but the determined time-scale is not a well-defined period.

\paragraph*{QY~Sge:} Menzies \& Whitelock (1988) reported low amplitude variations with
a characteristic time-scale of 50 days. Rao, Goswami \& Lambert (2002) interpreted their
spectra in terms of an obscured RV~Tauri variable. The NSVS data are quite noisy but show
four distinct minima of 0.08 mag at 60 days from each other.

\paragraph*{HD~213985:} Whitelock et al. (1989) observed a remarkably regular variability in
$K$-band with a period of 254 d and a peak-to-peak amplitude $\Delta K=0.29$ mag. Van Winckel,
Waelkens \& Waters (2000) measured an orbital period of 259 d from long-term radial velocity
measurements, implying that the light variations reflect orbital brightness modulation, either
from variable extinction or light scattering. The ASAS data confirm the period and show a
$V$-band amplitude ranging unpredictably  between 0.2--0.6 mag (Fig.\ \ref{lc3}), which means
the circumstellar shell must have variations on a similar time-scale to the orbital period.
The power spectrum does not contain any excess power that could indicate low-amplitude
pulsations.


\begin{thebibliography}{}

\bibitem[]{}
  Aikawa, T., 1991, ApJ, 374, 700
  
\bibitem[]{}
  Aikawa, T., 1993, MNRAS, 262, 893

\bibitem[]{}
  Alcock, C., et al., 1998, AJ, 115, 1921

\bibitem[]{}
  Baraffe, I., Alibert, Y., M\'era, D., Chabrier, G., Beaulieu, J.-P., 1998, ApJ, 499,
  L205

\bibitem[]{}
  Berdnikov, L.N., \& Turner, D.G., 2001, ApJS, 137, 209

\bibitem[]{}
  Bl\"ocker, T., \& Sch\"onberner, D., 1997, A\&A, 324, 991

\bibitem[]{}
  Bono, G., Marconi, M., Cassisi, S., Caputo, F., Gieren, W., \& Pietrzynski, G., 2005,
  ApJ, 621, 966

\bibitem[]{}
  Christensen-Dalsgaard, J., 2003, Lecture Notes on Stellar Oscillations, Aarhus
  University, May 2003, Fifth edition ({\tt 
  http://astro.phys.au.dk/$\sim$jcd/oscilnotes})
  
\bibitem[]{}
  De Ruyter, S., Van Winckel, H., Maas, T., Lloyd Evans, T., Waters, L.B.F.M., \& Dejonghe, H.,
  2006, A\&A, 448, 641

\bibitem[]{}
  Eggen, O.J., 1986, AJ, 91, 890
  
\bibitem[]{}
   ESA, 1997, The Hipparcos Catalogue, ESA SP-1200 

\bibitem[]{}
  Fokin, A.B., L\`ebre, A., Le Coroller, H., \& Gillet, D., 2001, A\&A, 378, 546

\bibitem[]{}
  Gautschy, A., 1993, MNRAS, 265, 340

\bibitem[]{}
  Giridhar, S., Rao, N.K., \& Lambert, D.L., 1992, JApA, 13, 307

\bibitem[]{}
  Giridhar, S., Rao, N.K., \& Lambert, D.L., 1994, ApJ, 437, 476

\bibitem[]{}
  Giridhar, S., Lambert, D.L., Bacham E., R., Gonzalez, G., Yong, D., 2005, ApJ, 627, 432

\bibitem[]{}
  Gonzalez, G., \& Wallerstein, G., 1996, MNRAS, 280, 515

\bibitem[]{}
  Harris, H.C., 1980, PhD Thesis, Univ. of Washington

\bibitem[]{}
  Hashimoto, O., 1994, A\&AS, 107, 445

\bibitem[]{}
  Hoffmeister, C., 1930, AN, 238, 17

\bibitem[]{}
  Hoffmeister, C., 1943, Kl. Ver\"off. Berlin-Babelsberg, No.\,27, 47

\bibitem[]{}
  Horowitz, D.H., 1986, JAAVSO, 15, 223

\bibitem[]{}
  Irwin, J.B., 1961, ApJS 6, 253

\bibitem[]{}
  Jurcsik, J., 1993, Acta Astron., 43, 353

\bibitem[]{}
  Kazarovets, E.V., Samus, N.N., \& Durlevich, O.V., 2001, IBVS, 5135

\bibitem[]{}
  Kholopov, P.N., et al., 1985-1988, General Catalogue of Variable Stars (GCVS), 4th edition,
  Nauka, Moscow

\bibitem[]{}
  Kilkenny D., et al., 1993, SAAO Circ., No.\,15, 85

\bibitem[]{}
  Lake, R., 1963, MNSSA, 22, 79

\bibitem[]{}
  Landolt, A.U., 1971, PASP, 83, 43

\bibitem[]{}
  Le Coroller, H., L\`ebre, A., Gillet, D., \& Chapellier, E., 2003, A\&A, 400, 613

\bibitem[]{}
  Lenz, P., \& Breger, M., 2005, Comm. Asteroseis., 146, 53

\bibitem[]{}
  Lloyd Evans, T., 1997, Ap\&SS, 251, 239

\bibitem[]{}
  Lloyd Evans, T., 1999, IAU Symp. 191, 453

\bibitem[]{}
  Maas, T., et al., 2003, A\&A, 405, 271

\bibitem[]{}
  Menzies, J.W., \& Whitelock, P.A., 1988, MNRAS, 233, 697

  
\bibitem[]{}
  Payne Gaposchkin, C., 1950, HA, 115, 205

\bibitem[]{}
  Percy, J.R., Molak, A., Lund, H., Overbeek, D., Wehlau, A.F., Williams, P.F., 2006,
  PASP, 118, 805

\bibitem[]{}
  Pojmanski, G., 2002, Acta Astron., 52, 397

\bibitem[]{}
  Pollard, K.R., \& Cottrell, P.L., 1995, ASP Conf. Series, 83, 409

\bibitem[]{}
  Pollard, K.R., Cottrell, P.L., Kilmartin, P.M., \& Gilmore, A.C., 1996, MNRAS, 279, 949
  
\bibitem[]{}
  Pollard, K.R., Cottrell, P.L., Lawson, W.A., Albrow, M.D., \& Tobin, W., 1997, MNRAS,
  286, 1 

\bibitem[]{}
  Rao, N.K., Goswami, A., \& Lambert, D.L., 2002, MNRAS, 334, 129

\bibitem[]{}
  Raveendran, A.V., 1999, MNRAS, 303, 595

\bibitem[]{}
  Saio, H., \& Gautschy, A., 1998, ApJ, 498, 360
 
\bibitem[]{}
  Soker, N., 2000, MNRAS, 312, 217

\bibitem[]{}
  Szczerba, R., G\'orny, S.K., \& Zalfresso-Jundzi\l\l o, M., 2001, Astrophys. Space Sci.
  Library, Vol. 265, 13

\bibitem[]{}
  Van Winckel, H., 2003, ARA\&A, 41, 391

\bibitem[]{}
  Van Winckel, H., Waelkens, C., \& Waters, L.B.F.M., 2000, IAU Symp., 177, 285

\bibitem[]{}
  Waelkens, C., et al., 1991, A\&A, 242, 433

\bibitem[]{}
  Waelkens, C., Van Winckel, H., Waters, L.B.F.M., \& Bakker, E.J., 1996, A\&A, 314, L17

\bibitem[]{}
  Wallerstein, G., 2002, PASP, 114, 689

\bibitem[]{}
  Walraven, Th. Muller, A.B., \& Oosterhoff, P.Th., 1958, BAN, 14, 81

\bibitem[]{}
  Whitelock, P.A., et al., 1989, MNRAS, 241, 393

\bibitem[]{}
  Wo\.zniak, P.R., et al., 2004, AJ, 127, 2436

\bibitem[]{}
  Zalewski, J., 1993, Acta Astron., 43, 431
    
\end{thebibliography}
\end{document}